\documentclass[aps,pre,reprint,letterpaper,showpacs,amsmath,amssymb]{revtex4-1}
\begin{document}
\title{Generalization of the Beck-Cohen superstatistics}
\received{16 September 2011}
\revised{24 October 2011}
\published{28 November 2011}
\author{Denis Nikolaevich \surname{Sob'yanin}}
\email{sobyanin@lpi.ru}
\affiliation{Tamm Department of Theoretical Physics,\\Lebedev Physical
Institute, Russian Academy of Sciences,\\Leninskii pr.\ 53, Moscow 119991,
Russia}
\begin{abstract}
Generalized superstatistics, i.e., a ``statistics of superstatistics,'' is
proposed. A generalized superstatistical system comprises a set of
superstatistical subsystems and represents a generalized hyperensemble. There
exists a random control parameter that determines both the density of energy
states and the distribution of the intensive parameter for each superstatistical
subsystem, thereby forming the third, upper level of dynamics. Generalized
superstatistics can be used for nonstationary nonequilibrium systems. The system
in which a supercritical multitype age-dependent branching process takes place
is an example of a nonstationary generalized superstatistical system. The theory
is applied to pair production in a neutron star magnetosphere.

\end{abstract}
\maketitle
\section{Introduction}
A large variety of complex nonequilibrium systems exhibit spatiotemporally
inhomogeneous dynamics. Such systems are often characterized by hierarchical
structures of dynamics. The hierarchy can be formed by the decomposition of the
system dynamics into different dynamics on different spatiotemporal scales,
which are largely separated from each other. In this case, the statistical
properties of the system can be effectively described by a superposition of
several statistics, or a ``superstatistics.''

Superstatistics has been formulated in Ref.~\cite{BeckCohen2003} to consider
nonequilibrium systems with a stationary state and intensive parameter
fluctuations. Some preliminary concepts have been anticipated earlier in
Refs.~\cite{WilkWlodarczyk2000,Beck2001,BashkirovSukhanov2002}. Superstatistical
systems are characterized by the existence of an intensive parameter $\beta$
that fluctuates on a much larger time scale than the typical relaxation time of
the local dynamics. If a given system can be thought of as a collection of many
small spatial cells, then the inverse temperature in a cell is typically taken
as such a parameter. However, more general interpretations of the intensive
parameter are possible. Sufficient time scale separation between two relevant
dynamics within the complex system allows one to qualify superstatistics as a
form of slow modulation \cite{AllegriniEtal2006}.

Superstatistics is applicable to various complex systems. Its applications
include, among others, cosmic-ray energy spectra and electron-positron pair
annihilation \cite{Beck2004,Beck2009}, the world line representations of Feynman
propagators for spin-$0$ and spin-$1/2$ particles \cite{JizbaKleinert2010}, an
extension of the random matrix theory covering systems with mixed
regular-chaotic dynamics \cite{AbulMagd2006a,AbulMagd2006b,AbulMagdEtal2008},
nonstationary dynamical processes with time-varying multiplicative noise
exponents \cite{DuarteQueiros2008}, Markovian systems without detailed balance
\cite{LubashevskyEtal2009}, a mesoscopic approach to the problem of Brownian
motion \cite{RodriguezSantamariaHolek2007}, models of the metastatic cascade in
cancerous systems \cite{ChenBeck2008}, complex networks \cite{AbeThurner2005},
ecosystems driven by hydroclimatic fluctuations \cite{PorporatoVicoFay2006},
pattern-forming systems \cite{DanielsBeckBodenschatz2004}, solar flares
\cite{BaiesiPaczuskiStella2006}, share price fluctuations
\cite{AusloosIvanova2003,JizbaKleinert2008,AnteneodoDuarteQueiros2009,
StraetenBeck2009}, the statistics of train departure delays
\cite{BriggsBeck2007}, wind velocity fluctuations \cite{RizzoRapisarda2004}, and
many interesting applications in hydrodynamic turbulence
\cite{Reynolds2003,JungSwinney2005,BeckCohenSwinney2005,Beck2007,
StraetenBeck2009,Abe2010}.

Let us suppose for a while that we have a complex nonequilibrium system
described by superstatistics. One of the main problems is the determination of
the distribution $f(\beta)$ of the intensive parameter $\beta$. It can be found
using the maximum entropy principle \cite{AbeBeckCohen2007}. Moreover,
$f(\beta)$ can be considered as a function of a set of additional control
parameters $\{\lambda_i\}$, $f=f(\beta,\{\lambda_i\})$ \cite{StraetenBeck2008}.
These parameters emerge as Lagrange multipliers determined by maximizing the
Boltzmann-Gibbs-Shannon entropy of $f(\beta,\{\lambda_i\})$ under certain
constraints. Let us revert to the parameter $\beta$. In each cell of the
superstatistical system, we have the Gibbs canonical distribution
$\rho_G(E|\beta)=e^{-\beta E}/Z(\beta)$, where $Z(\beta)$ is the partition
function. This distribution can be obtained by maximizing the
Boltzmann-Gibbs-Shannon entropy associated with a normalized distribution
$\rho(E|\beta)$, where $\beta$ is the Lagrange multiplier corresponding to the
mean energy constraint. We see that, in general, both $\beta$ and
$\{\lambda_i\}$ have a similar nature. However, in superstatistics, $\beta$
fluctuates, but the control parameters are constant.

We can pose the following question: What statistics appears in the case of
fluctuating control parameters? The aim of this paper is to develop the
generalization of the Beck-Cohen superstatistics that allows one to properly
consider such fluctuations. These form a separate, upper level of dynamics, thus
leading to multiscale superstatistics. The possible existence of such
superstatistics has been presumed in Ref.~\cite{AbeBeckCohen2007}.

This paper is organized as follows. In Sec.~II generalized superstatistics is
developed. In Sec.~III supercritical multitype age-dependent branching processes
are considered as an example of generalized superstatistical systems. In Sec.~IV
generalized superstatistics is applied to electron-positron pair production in a
neutron star magnetosphere and the particle energy distribution is calculated.
In Sec.~V the main results are summarized.

\section{Generalized superstatistics}

Let us consider a nonequilibrium system that consists of a set of nonequilibrium
subsystems described by their superstatistics. This system can be referred to as
a ``generalized superstatistical system.'' Let us define a random variable $\xi$
that determines the properties of each nonequilibrium superstatistical
subsystem. The role of $\xi$ in determining the properties of the subsystems is
similar to that of $\beta$ in determining the properties of the cells that
constitute a superstatistical subsystem. In other words, $\xi$ is a control
parameter, which determines the form of the superstatistical distribution for
each subsystem. Here, we do not restrict ourselves to considering $\xi$ as a
scalar random variable. In principle, $\xi$ is allowed to be a multidimensional
random vector. The distribution of $\xi$ is characterized by a probability
density $c(\xi)$, which is normalized, $\int c(\xi)d\xi=1$.

We can look at the generalized superstatistical system using the hyperensemble
approach \cite{Crooks2007}. Developed in the context of nonequilibrium systems,
this approach is consistent with the standard theory of equilibrium statistical
physics \cite{Naudts2007}. Superstatistics can be considered as the theory of
hyperensembles \cite{Abe2009}. Each superstatistical subsystem of the system
represents a hyperensemble, i.e., a mixture of canonical ensembles. In turn, the
system as a whole can be described by a mixture of the hyperensembles
corresponding to the subsystems. In other words, the generalized
superstatistical system represents a ``generalized hyperensemble,'' i.e., an
ensemble of hyperensembles.

There are three levels of dynamics in the system under consideration: The first
level is the level of fast dynamics in a cell. The second level is the level of
superstatistical dynamics in a subsystem. The existence of an additional
dynamics described by $\xi$ forms the third level in the hierarchy of dynamics
and allows us to refer to the statistics of the generalized superstatistical
system as a ``generalized superstatistics,'' i.e., a statistics of
superstatistics.

Consider a superstatistical subsystem of the generalized superstatistical
system. Let $\Gamma(E|\xi)$ be a nondecreasing function representing the number
of states with energy less than $E$. The random variable $\xi$ determines the
density of energy states for the subsystem,
\[
g(E|\xi)=\frac{\partial\Gamma(E|\xi)}{\partial E}.
\]
The inclusion of the density of states in the superstatistical description is
significant \cite{Sattin2006}. When writing $d\Gamma(E|\xi)$ in integrals, we
will imply integration over $E$ so that $d\Gamma(E|\xi)=g(E|\xi)dE$. The Gibbs
canonical distribution for each cell of the subsystem is
\begin{equation}
\label{GibbsDistributionWithXi}
\rho_G(E|\beta,\xi)=\frac{e^{-\beta E}}{Z(\beta|\xi)},
\end{equation}
where
\[
Z(\beta|\xi)=\int e^{-\beta E}d\Gamma(E|\xi)
\]
is the partition function. By analogy with the approach used in
Ref.~\cite{StraetenBeck2008}, we do not include the density of states in the
definition of the energy distribution \eqref{GibbsDistributionWithXi};
therefore, the normalization condition is
$\int\rho_G(E|\beta,\xi)d\Gamma(E|\xi)=1$. We also assume that $\xi$ determines
the distribution $f(\beta|\xi)$ of the intensive parameter $\beta$. This
distribution is normalized, $\int f(\beta|\xi)d\beta=1$. Note that $g(E|\xi)$
and $f(\beta|\xi)$ are not necessarily statistically dependent variables. These
will change independently if we choose $\xi=(\xi_1,\xi_2)$, with $\xi_1$ and
$\xi_2$ being independent random variables, and set $g(E|\xi)=g(E|\xi_1)$ and
$f(\beta|\xi)=f(\beta|\xi_2)$.

The superstatistical distribution for the subsystem is given by
\begin{equation}
\label{superstatisticalDistributionWithXi}
\rho(E|\xi)=\int\rho_G(E|\beta,\xi)f(\beta|\xi)d\beta,
\end{equation}
with the normalization condition being $\int\rho(E|\xi)d\Gamma(E|\xi)=1$. If
$\xi$ is a nonrandom vector, then this distribution reduces to the ordinary
superstatistical distribution \cite{BeckCohen2003}. By averaging
Eq.~\eqref{superstatisticalDistributionWithXi} over the fluctuating $\xi$, we
immediately obtain the generalized superstatistical distribution
\begin{equation}
\label{generalizedSuperstatisticalDistribution}
\sigma(E)=\int\rho(E|\xi)g(E|\xi)c(\xi)d\xi,
\end{equation}
which is normalized, $\int\sigma(E)dE=1$.

\section{Branching processes}
Now the following question arises: Do there exist any nonequilibrium systems,
whether stationary or not, that can be described by generalized superstatistics?

Let us consider a many-particle system composed of particles of $n$ types. Each
type-$i$ particle, which will also be denoted by $T_i$, $1\leqslant i\leqslant
n$, has a random lifetime with a probability distribution function $G_i(\tau)$.
In other words, $G_i(\tau)$ is the probability that the lifetime of a given
type-$i$ particle does not exceed $\tau$. At the end of its life the particle
decays into a random number of particles of several types.
Specifically, at the moment of its decay the particle produces
$\omega_j\geqslant0$ type-$j$ particles of age zero, $1\leqslant j\leqslant n$:
\begin{equation}
\label{TiTransformation}
T_i\rightarrow\sum_{j=1}^n\omega_j T_j.
\end{equation}
Thus we have a multitype age-dependent branching process, the so-called
multitype Sevast'yanov process \cite{Sevastyanov1964}.

The transformation \eqref{TiTransformation} is described by the generating
function
\[
h_i(\tau,x)=\sum_\omega p_i^\omega(\tau)x_1^{\omega_1}\cdots x_n^{\omega_n},
\]
where $p_i^\omega(\tau)$ is the conditional probability of the transformation
\eqref{TiTransformation} given that the particle decays at age $\tau$,
$\omega=(\omega_1,\ldots,\omega_n)$ is an $n$-dimensional vector with nonnegative
integer components $\omega_j$, and $x=(x_1,\ldots,x_n)$ is an $n$-dimensional
vector with complex components $x_j$ such that $|x_j|\leqslant1$. If
$x_1=\cdots=x_n=1$, we will write $x=1$. The normalization condition reads
$h_i(\tau,1)$=1.

The mean number of type-$j$ particles that appear upon the decay of a type-$i$
particle is given by
\begin{equation}
\label{matrixA}
A_{ij}=\int_0^\infty a_{ij}(\tau)dG_i(\tau),
\end{equation}
where
\[
a_{ij}(\tau)=\left.\frac{\partial h_i(\tau,x)}{\partial x_j}\right|_{x=1}
\]
is the same mean given that the particle decays at age $\tau$. We will assume that
the $n\times n$ matrix $A=\|A_{ij}\|$ with components $0\leqslant A_{ij}<\infty$
is irreducible, or indecomposable, i.e., the index set $\{1,\ldots,n\}$ cannot be
divided into two disjoint nonempty sets $S_1$ and $S_2$ such that $A_{ij}=0$ for
all $i\in S_1$ and all $j\in S_2$ (see, e.g., Ref.~\cite{Gantmacher1959} for
details). Moreover, we will also assume that the Perron root of $A$, i.e., the
maximum positive real eigenvalue of $A$, is greater than one. Thus we deal with
the indecomposable supercritical branching process \cite{VatutinZubkov1985}.
Physically, this means that, first, a particle of a given type potentially has
descendants, either direct or distant, of any type and, second, the number of
particles in the system, on average, progressively increases.

We see that the whole system is composed of $n$ subsystems, the $i$th subsystem
comprising type-$i$ particles. Obviously, the subsystems interact with each
other in the sense that the decay of a particle in one subsystem leads to the
creation of particles in other subsystems. The number of particles both in the
whole system and in the subsystems is not constant and changes with time. Thus
we have a nonstationary nonequilibrium situation.

Let $\mu(t)=[\mu_1(t),\ldots,\mu_n(t)]$ be an $n$-dimensional vector with
non-negative integer components. Each $\mu_j(t)$ is a random variable that yields
the number of particles in the $j$th subsystem at time $t$. The age-dependent
branching process is characterized by a set of generating functions
\begin{equation}
\label{FGeneratingFunction}
F_i(t,x)=\sum_\omega P_i^\omega(t)x_1^{\omega_1}\cdots x_n^{\omega_n},
\end{equation}
where $P_i^\omega(t)$ is the conditional probability that $\mu(t)=\omega$ given
that there is one particle in the $i$th subsystem and none in the other
subsystems at time zero. The generating functions $F_i(t,x)$
\eqref{FGeneratingFunction}, $1\leqslant i\leqslant n$, satisfy a system of
nonlinear integral equations \cite{Sevastyanov1964}. Essential for us here is
the probabilistic nature of the number of particles both in the whole system and
in each subsystem. In general, we cannot predict \textit{a priori} what the number of
particles is in each subsystem at a given time, but instead we can focus on the
average behavior of the generalized hyperensemble.

The mean number of particles in the $j$th subsystem at time $t$ given that the
branching process has started with one type-$i$ particle at time zero is
\[
A_{ij}(t)=\left.\frac{\partial F_i(t,x)}{\partial x_j}\right|_{x=1}.
\]
To analyze the long-run behavior of these means, let us define the
Laplace-Stieltjes transforms
\[
L_{ij}(\alpha)=\int_0^\infty e^{-\alpha\tau}dG_{ij}(\tau),
\]
where
\begin{equation}
\label{Gij}
G_{ij}(t)=\int_0^t a_{ij}(\tau)dG_i(\tau).
\end{equation}
Let $\lambda(\alpha)$ be the Perron root of the $n\times n$ matrix
$L(\alpha)=\|L_{ij}(\alpha)\|$. Choose an $\alpha$ such that $\lambda(\alpha)=1$
and define an $n$-dimensional left eigenvector $v=(v_1,\ldots,v_n)$ of $L(\alpha)$
with positive real components $v_j$ that satisfy
\[
v_j=\sum_{i=1}^n v_i L_{ij}(\alpha).
\]
Then the asymptotic behavior of the mean numbers of particles is expressed as
\begin{equation}
\label{AijT}
A_{ij}(t)\sim C_i v_j w_j e^{\alpha t},\qquad t\rightarrow\infty,
\end{equation}
where
\begin{equation}
\label{wj}
w_j=\int_0^\infty e^{-\alpha\tau}[1-G_j(\tau)]d\tau
\end{equation}
and $C_i$ is a positive constant \cite{SevastyanovChistyakov1970}.
Interestingly, $A_{ij}(t)$ behaves in a similar way for each subsystem. Since we
consider the supercritical case, $\alpha$ always exists and is positive.
Therefore, the mean number of particles in each subsystem increases
exponentially. Note that the existence of a positive $\alpha$
is equivalent to the condition that the Perron root of $A$
is greater than one \cite{VatutinZubkov1985,SevastyanovChistyakov1970}.

In the long-term run, the limiting probability that a given particle belongs to
the $i$th subsystem can be written as
\begin{equation}
\label{limitingProbability}
\pi_i=\frac{v_i w_i}{\sum_{j=1}^n v_j w_j}.
\end{equation}
Note that $\pi_i$ is independent of the type of the primary particle with which
the branching process has started. Moreover, nonstationary though the situation
is, the limiting probability is stationary.

The limiting age distribution for the $i$th subsystem is
\begin{equation}
\label{ageDistribution}
L_i(\tau)=\frac{\int_0^\tau e^{-\alpha u}[1-G_i(u)]du}{\int_0^\infty
e^{-\alpha u}[1-G_i(u)]du}.
\end{equation}
It is found by dividing the $i$th subsystem into two subsystems, the first
comprising the type-$i$ particles that decay at age $u\leqslant\tau$ and the
second comprising the rest, and applying the technique described above to the
new system. Equation~\eqref{ageDistribution} yields the probability that a
randomly chosen type-$i$ particle decays at age $u\leqslant\tau$. In the case of a
single type of particles, $L_i(\tau)$ reduces to the limiting age distribution
for the one-dimensional supercritical Sevast'yanov process
\cite{YakovlevYanev2007}, which coincides with the classical age distribution
for the one-dimensional supercritical Bellman-Harris process when the
probability $p_i^\omega(\tau)$ is independent of $\tau$,
$p_i^\omega(\tau)=p_i^\omega$ \cite{AthreyaKaplan1976,Kuczek1982}.

Now we can calculate the particle energy distribution. Consider a type-$i$
particle of age $\tau$. In general, its energy can be considered as a random
variable characterized by a conditional probability density $w_i(E|\tau)$. More
specifically, the probability that the energy of a type-$i$ particle of age
$\tau$ lies in a small interval $dE$ around $E$ is $w_i(E|\tau)d\Gamma_i(E)$,
where $\Gamma_i(E)$ is the number of energy states with energy less than $E$.
The normalization condition is $\int w_i(E|\tau)d\Gamma_i(E)=1$. The energy
probability density for the $i$th subsystem becomes
\begin{equation}
\label{energyDistributionForSubsystem}
\rho_i(E)=\int_0^\infty w_i(E|\tau)dL_i(\tau),
\end{equation}
with the normalization condition being $\int\rho_i(E)d\Gamma_i(E)=1$.

The described system can be considered as a generalized superstatistical system.
The control parameter $\xi$ is a discrete random variable that takes on values
$\{1,\ldots,n\}$ and yields the number of the subsystem to which a randomly chosen
particle belongs. In other words, $\xi$ corresponds to the particle type.
It has the discrete probability distribution
$\{\pi_1,\ldots,\pi_n\}$, where $\pi_i$ is given by
Eq.~\eqref{limitingProbability}. It remains to find the distribution
$f_i(\beta)$ of the fluctuating parameter $\beta$ for each subsystem. Let
\[
g(s)=\mathfrak{L}[f(x)](s)=\int_0^\infty e^{-sx}f(x)\,dx
\]
be the Laplace transform of a function $f(x)$, with
$f(x)=\mathfrak{L}^{-1}[g(s)](x)$ being the corresponding inverse Laplace
transform. The energy distribution for the $i$th subsystem can be expressed as
\[
\rho_i(E)=\mathfrak{L}\biggl[\frac{f_i(\beta)}{Z_i(\beta)}\biggr](E),
\]
where $Z_i(\beta)$ is the partition function. Then
\[
f_i(\beta)=Z_i(\beta)\mathfrak{L}^{-1}[\rho_i(E)](\beta).
\]

\section{An example: pair production in a neutron star magnetosphere}

New nonstationary cosmic radio sources associated with neutron stars, viz.,
intermittent pulsars \cite{KramerEtal2006} and rotating radio transients (RRATs)
\cite{McLaughlinEtal2006}, have been discovered recently. The essential feature
of these sources is their long ``silence,'' when we do not observe any radio
emission from them. Since an electron-positron plasma outflowing from the
magnetosphere of a neutron star is responsible for the observable radio
emission, the plasma generation can be switched off for some time. In this case,
the absorption of a high-energy photon in the inner neutron star magnetosphere
triggers nonstationary cascade pair production \cite{IstominSobyanin2011a},
which, in turn, results in the formation of a ``lightning''
\cite{IstominSobyanin2011b}. Such lightnings can manifest themselves as radio
bursts from RRATs \cite{IstominSobyanin2011c}. In a lightning, we deal with an
ultrarelativistic electron-positron plasma. The properties of the emission from
electrons and positrons are determined by their energies. Therefore, it is
important to find the energy distribution of particles. This can be done using
generalized superstatistics.

Let us characterize the energy of a charged particle, either an electron or a
positron, by its Lorentz factor $\gamma(\tau)$. The particle is efficiently
accelerated by a longitudinal electric field $E_\parallel$ so that
$\gamma(\tau)$ eventually reaches a stationary value $\gamma_0$, which is
$\sim\!\!10^8$ in a vacuum neutron star magnetosphere \cite{IstominSobyanin2009}.
The electron and positron of each produced pair, though ultrarelativistic,
initially have Lorentz factors much less than $\gamma_0$. At the initial stage
of acceleration $\gamma(\tau)$ increases linearly with time,
$\gamma(\tau)\approx E_\parallel \tau$. Here, we use a dimensionless system of
units (see, e.g., Ref.~\cite{IstominSobyanin2011c}). By contrast, when $t$ approaches
$\tau_0=\gamma_0/E_\parallel$, the radiation forces come to the fore, and a need
arises to use the Dirac-Lorentz equation to consider the particle dynamics
properly \cite{IstominSobyanin2009,IstominSobyanin2010a,IstominSobyanin2010b}.

Thus, we can define two types of particles: A type-$1$ particle can be
efficiently accelerated by the electric field since the radiation friction is
negligible. A type-$2$ particle, in contrast to a type-$1$ particle, is not
accelerated by the electric field because of the electrodynamic self-action
effects and has the constant Lorentz factor $\gamma_0$. In the former case the
particle does not efficiently produce secondary pairs, but in the latter case it
does at a rate $Q$ \cite{IstominSobyanin2011a}. Note that the plasma generation,
along with the accompanying radio emission, is not suppressed even in ultrahigh
magnetar magnetic fields \cite{IstominSobyanin2007,IstominSobyanin2008}. The
Lorentz factors of type-$1$ and type-$2$ particles as functions of their ages
become
\begin{subequations}
\label{LorentzFactors}
\begin{alignat}{2}
\gamma_1(\tau)&=E_\parallel \tau,\qquad&0&\leqslant\tau<\tau_0,\\
\gamma_2(\tau)&=\gamma_0,\qquad&0&\leqslant\tau<\infty.
\end{alignat}
\end{subequations}

The particles of each produced pair, though moving independently of each other,
can conveniently be considered as a whole, and type-$1$ and type-$2$ pairs,
which will be denoted by $T_1$ and $T_2$, respectively, are defined by analogy
with individual particles.

Now we can write the following transformations:
\begin{eqnarray*}
T_1&\rightarrow&T_2,\label{T1}\\
T_2&\rightarrow&T_1+T_2.\label{T2}
\end{eqnarray*}
The generating functions are
\begin{eqnarray*}
h_1(\tau,x)&=&x_2,\\
h_2(\tau,x)&=&x_1 x_2,
\end{eqnarray*}
where $x=(x_1,x_2)$. The $2\times2$ matrix $a=\|a_{ij}\|$ becomes
$\bigl(
\begin{smallmatrix}
0&1\\
1&1
\end{smallmatrix}
\bigr)$.
The lifetime distribution functions are
\begin{subequations}
\label{ageDistributions}
\begin{eqnarray}
G_1(\tau)&=&\theta(\tau-\tau_0),\\
G_2(\tau)&=&1-e^{-2Q\tau},
\end{eqnarray}
\end{subequations}
where $\theta(x)$ is the Heaviside function. The matrix $A$ defined by Eq.~\eqref{matrixA} coincides with $a$
and is indecomposable since its off-diagonal elements are positive.
The Perron root of $A$ is equal to $(1+\sqrt{5})/2$ and greater than~$1$.
Therefore, we deal with the indecomposable supercritical two-type branching process.

Now we should find a proper value of $\alpha$ as described after
Eq.~\eqref{Gij}. Interestingly, we can do this without directly finding the
Perron root of $L(\alpha)$. We have recently shown \cite{IstominSobyanin2011a}
that pair production in the system under consideration is asymptotically
described by the equation
\begin{equation}
\label{effectiveEquation}
\frac{dN(t)}{dt}=2Q^{\mathrm{eff}}N(t),
\end{equation}
where $N(t)$ is the number of electron-positron pairs at time $t$,
$Q^{\mathrm{eff}}=N^{\mathrm{eff}}_{\tau_0}/2\tau_0$ is the effective pair production rate, and
$N^{\mathrm{eff}}_{\tau_0}$ satisfies
\[
N^{\mathrm{eff}}_{\tau_0}=\ln N_{\tau_0}-\ln N^{\mathrm{eff}}_{\tau_0},
\]
where $N_{\tau_0}=2Q\tau_0$ is the number of particles created by a fully
accelerated particle in time $\tau_0$. Note that $N^{\mathrm{eff}}_{\tau_0}$ can be
equivalently expressed through the Lambert function $W(x)$ (see, e.g.,
Ref.~\cite{AsgaraniMirza2008} for its definition) as
\[
N^{\mathrm{eff}}_{\tau_0}=W(N_{\tau_0})
\]
and hence is positive since $W(x)$ is positive for any positive $x$.
From Eq.~\eqref{effectiveEquation} it follows that the total number of
electron-positron pairs in the system increases exponentially. Comparing
Eq.~\eqref{effectiveEquation} to Eq.~\eqref{AijT} yields
\begin{equation}
\label{alpha2Qeff}
\alpha=2Q^{\mathrm{eff}},
\end{equation}
which is positive, as it must be in the supercritical case.

The energy of any type-$1$ particle is less than $\gamma_0$, while that of any
type-$2$ particle is $\gamma_0$. Therefore, it is natural to choose the following
density of states for the subsystems:
\begin{subequations}
\label{g1andg2}
\begin{eqnarray}
g_1(\gamma)&=&1-\theta(\gamma-\gamma_0),\\
g_2(\gamma)&=&\delta(\gamma-\gamma_0),
\end{eqnarray}
\end{subequations}
where $\delta(x)$ is the delta function. We see from Eq.~\eqref{LorentzFactors}
that the energies of particles are nonrandom functions of their ages, hence the
conditional energy distributions have the form
\begin{subequations}
\label{energyDistributionsGivenTau}
\begin{eqnarray}
w_1(\gamma|\tau)&=&\delta[\gamma-\gamma_1(\tau)],\\
w_2(\gamma|\tau)&=&1.
\end{eqnarray}
\end{subequations}
Using Eqs.~\eqref{ageDistribution}, \eqref{energyDistributionForSubsystem},
\eqref{ageDistributions}, and \eqref{energyDistributionsGivenTau}, we obtain the
energy distributions for the subsystems:
\begin{subequations}
\label{unconditionalEnergyDistributions}
\begin{eqnarray}
\rho_1(\gamma)&=&\frac{\alpha}{E_\parallel}\frac{
e^{-\alpha\gamma/E_\parallel}}{1-e^{-\alpha\tau_0}},\\
\rho_2(\gamma)&=&1.
\end{eqnarray}
\end{subequations}
The corresponding intensive parameter distributions are
\begin{eqnarray*}
f_1(\beta)&=&\delta\Bigl(\beta-\frac{\alpha}{E_\parallel}\Bigr),\\
f_2(\beta)&=&\delta(\beta).
\end{eqnarray*}

Let us find the probability distribution of the random control parameter $\xi$
that corresponds to the type of a randomly chosen particle.
From Eqs.~\eqref{wj} and \eqref{ageDistributions} we have
\begin{subequations}
\label{w1andw2}
\begin{eqnarray}
w_1&=&\frac{1-e^{-\alpha\tau_0}}{\alpha},\\
w_2&=&\frac{1}{\alpha+2Q}.
\end{eqnarray}
\end{subequations}
The components of the left eigenvector $v=(v_1,v_2)$ of
\[
L(\alpha)=%
\begin{pmatrix}
0& e^{-\alpha\tau_0}\\
\bigl(1+\frac{\alpha}{2Q}\bigr)^{-1}&\bigr(1+\frac{\alpha}{2Q}\bigl)^{-1}
\end{pmatrix}
\]
can be chosen as follows:
\begin{subequations}
\label{v1andv2}
\begin{eqnarray}
v_1&=&1,\\
v_2&=&1+\frac{\alpha}{2Q}.
\end{eqnarray}
\end{subequations}
Using Eqs.~\eqref{limitingProbability}, \eqref{alpha2Qeff}, \eqref{w1andw2}, and
\eqref{v1andv2}, we obtain the probability $\pi_\xi$ that a randomly chosen
particle is of type~$\xi$, $\xi=1,2$:
\begin{subequations}
\label{p1andp2}
\begin{eqnarray}
\pi_1&=&1-\frac{\alpha}{2Q},\\
\pi_2&=&\frac{\alpha}{2Q}.
\end{eqnarray}
\end{subequations}
Note that $\pi_2$ may be interpreted as the probability that the particle
significantly contributes to pair production.

Finally, Eqs.~\eqref{generalizedSuperstatisticalDistribution},
\eqref{alpha2Qeff}, \eqref{g1andg2}, \eqref{unconditionalEnergyDistributions},
and \eqref{p1andp2} allow us to obtain the generalized superstatistical
distribution
\begin{equation}
\label{particleEnergyDistribution}
\sigma(\gamma)=
\frac{\alpha}{2Q}\,\delta(\gamma-\gamma_0)
+[1-\theta(\gamma-\gamma_0)]\frac{\alpha}{E_\parallel}e^{
-\alpha\gamma/E_\parallel},
\end{equation}
which represents the energy distribution of ultrarelativistic electrons and
positrons.

The particle energy distribution in the neutron star magnetosphere cannot be
observed directly. Moreover, it is unclear how to infer the particle energy
distribution for RRATs from currently available observational data. On the one
hand, it is difficult to detect radio emission from RRATs because of its
sporadic nature, and one has to carry out single-pulse searches, which require
significant radio telescope resources with long observation times
\cite{McLaughlinEtal2009,KeaneEtal2011}. On the other hand, it is unclear how
the form of the distribution correlates with the properties of radio emission.
Although the parameters of an electron-positron plasma generated in the RRAT
magnetosphere imply the possibility of the efficient generation of radio emission
\cite{IstominSobyanin2011b,IstominSobyanin2011c}, the detailed mechanism is
unknown at the moment. Therefore, the distribution
\eqref{particleEnergyDistribution} is a theoretical prediction.

\section{Conclusion}
I have developed generalized superstatistics, i.e., a statistics of
superstatistics. Based on the concept of fluctuating control parameters, it is
defined for a generalized superstatistical system, which consists of a set of
nonequilibrium superstatistical subsystems. Such a system can be considered as a
generalized hyperensemble, i.e., an ensemble of hyperensembles. A fluctuating
control parameter, which may be a multidimensional random vector, determines the
form of the superstatistical distribution for each subsystem. In general, it
determines not only the distribution of the intensive parameter but also the
density of energy states. In a generalized superstatistical system, there
appears the third, upper level in the hierarchy of dynamics besides two levels
that exist in each superstatistical subsystem. Interestingly, generalized
superstatistics can be applied to nonstationary nonequilibrium systems. As an
example of a nonstationary generalized superstatistical system, I have studied
the system in which the supercritical multitype Sevast'yanov process takes place.
In this system, the transformation of particles of several types
occurs, and the number of particles both in the whole system and in each
subsystem increases exponentially. In addition, I have considered an
astrophysical application of generalized superstatistics and obtained the energy
distribution of ultrarelativistic electrons and positrons produced in a neutron
star magnetosphere. I have found the probability that a randomly chosen particle
significantly contributes to the production of secondary electron-positron
pairs.

It seems interesting to find general principles that might allow us to obtain
the distribution of both the intensive parameter and the control parameter. Note
that in the case of ordinary superstatistics, the main approach to finding the
intensive parameter distribution is maximizing either a generalized entropy or
the Boltzmann-Gibbs-Shannon entropy under certain constraints
\cite{TsallisSouza2003,AbeBeckCohen2007,Crooks2007,Naudts2007,StraetenBeck2008,
Abe2009,Abe2010}. I believe that the same approach might be used in the case of
generalized superstatistics. However, this problem requires a separate study.

\providecommand{\noopsort}[1]{}\providecommand{\singleletter}[1]{#1}%
\end{document}